# Out of plane effect on the superconductivity of $Sr_{2-x}Ba_xCuO_{3+\delta}$ with Tc up to 98K

W. B. Gao, Q. Q. Liu, L. X. Yang, Y. Yu, F. Y. Li, C. Q. Jin*

*Institute of Physics, Chinese Academy of Sciences, Beijing, China*

S. Uchida*

*Department of Physics, University of Tokyo, 7-3-1 Hongo, Bunkyo-ku, Tokyo 113-0033, Japan*

## Abstract

A series of new $Sr_{2-x}Ba_xCuO_{3+\delta}$ ($0 \leq x \leq 0.6$) superconductors were prepared using high-pressure and high-temperature synthesis. A Rietveld refinement based on powder x-ray diffraction confirms that the superconductors crystallize in the $K_2NiF_4$-type structure of a space group I4/mmm similar to that of $La_2CuO_4$ but with partially occupied apical oxygen sites. It is found that the superconducting transition temperature $T_c$ of this Ba substituted $Sr_2CuO_{3+\delta}$ superconductor with constant carrier doping level, *i.e.*, constant δ, is controlled not only by order/disorder of apical-O atoms but also by Ba content. $T_c^{max}$ =98 K is achieved in the material with x=0.6 that reaches the record value of $T_c$ among the single-layer copper oxide superconductors, and is higher than $T_c$=95K of $Sr_2CuO_{3+\delta}$ with optimally ordered apical-O atoms. There is Sr-site disorder in $Sr_{2-x}Ba_xCuO_{3+\delta}$ which might lead to a reduction of $T_c$. The result indicates that another effect surpasses the disorder effect that is related either to the increased in-plane Cu-O bond length or to elongated apical-O distance due to Ba substitution with larger cation size. The present experiment demonstrates that the optimization of local geometry out of the Cu-O plane can dramatically enhance $T_c$ in the cuprate superconductors.

**Correspondance** authors are C. Q. Jin (JIN@aphy.iphy.ac.cn) & S. Uchida (UCHIDA@phys.s.u-tokyo.ac.jp)

# 1. Introduction

The enhancement of the superconducting transition temperature ($T_c$) is one of key concerns for the studies of high temperature superconductors (HTS). Crystallographically HTS consists of $CuO_2$ plane that supports superconducting current, as well as charge reservoir layer where doping takes place. In the hole-doped high-Tc cuprates the charge reservoir is composed of two parts. That immediately outside the $CuO_2$ plane is called as the apical oxygen (apical-O) layer since the oxygen atoms here are at the apical sites of a $CuO_6$ octahedron or a $CuO_5$ pyramid, and that outside the apical-O layer, is named the 2$^{nd}$ nearest charge reservoir layer. The cation in the apical-O layer is termed as the A-site cation. The apical-O and A-site cation are expected to be more heavily involved in the electronic state of the $CuO_2$ plane than the 2$^{nd}$ nearest charge reservoir layer via interaction with $CuO_2$ plane since they have direct chemical bonding.

There seem to be several factors to "optimize" the electronic structures of $CuO_2$ plane in order to achieve a higher $T_c$. First of all, to induce HTS, one has to dope the parent Mott compounds, i.e. generate carriers into $CuO_2$ plane. Consequently, dopant concentration, or carrier density $n_h$ is the major factor controlling $T_c$. It has been found that $T_c$ generally follows a parabolic relation as function of $n_h$ showing a maximum at an optimum doping $n_h^{opt}$.[1]

However, the maximum $T_c$ varies widely among the known materials that share the same $CuO_2$ plane. Particularly, in the monolayer cuprates $T_c$ differs by a factor of 3.[2] Therefore, there must be significant effects on $T_c$ from the charge-reservoir layers

(out-of-plane effects). Primary doping mechanism for p-type HTS is chemical substitution with dopant metal at A-site that has a different valence from the host ion. For example, Sr is a dopant for $(La,Sr)_2CuO_4$ superconductor. While doping changes carrier density in the $CuO_2$ plane, the chemical substitution unavoidably introduces disorder into the crystalline lattice due to random distribution of dopant atoms. Thus, the dopant disorder/order is a primary out-of-plane effect that influences $T_c$. The effect of dopant disorder on $T_c$ has become a recent concern. The effect of the out-of-plane disorder was systematically studied for $La_2CuO_4$,[3] or Nd- and Bi-based copper oxide superconductors.[4] It was demonstrated that the introduction of cation disorder in the apical-O layer, A-site disorder, gives rise to an appreciable decrease in $T_c$.[2,3,4] The scanning-tunneling spectroscopy (STS) has revealed variations in the measured gap on nanometer scale. This gap inhomogeneity is shown also by STS to be correlated with the out-of-plane disorder.[5-11]

Tuning oxygen stoichiometry is another usual way to dope $CuO_2$ plane as typically found for YBCO 123, Bi 22(n-1)n, or Hg12(n-1)n where $n_h$ is controlled by the oxygen amount in the 2$^{nd}$ nearest charge reservoir layer. Since there is a direct connection with $CuO_2$ plane, tuning apical oxygen is expected to have stronger effect than those at the 2$^{nd}$ charge reservoir layer on superconductivity. This was supported by recent theoretical work that indicates correlation of $T_c$ with the distance between apical oxygen and $CuO_2$ plane.[12] However, for almost all the hole-doped HTS the apical oxygen sites are fully occupied. Searching for a system with tunable apical oxygen content would be effective for studying the superconducting mechanism of

HTS. The hole doped $Sr_2CuO_{3+\delta}$ superconductor synthesized under high pressure is a unique cuprate that crystallizes into an oxygen-deficient $La_2CuO_4$ (*i.e.* $K_2NiF_4$) structure with partially occupied apical sites as shown in Fig. 1. Interestingly the apical oxygen acts as hole dopant,[13,14,15] i.e. the carrier density of $Sr_2CuO_{3+\delta}$ is determined by the amount of additional oxygen $\delta$ that is located at the apical site. Recently, using $SrO_2$ as an oxidizer, we have succeeded in fabricating the tetragonal single-phase $Sr_2CuO_{3+\delta}$ that was free of Cl containing phase or other contamination claimed in previous reports.[15] Our studies show that the maximum $T_c$ can be achieved for $Sr_2CuO_{3+\delta}$ with a nominal doping level at $\delta=0.4$ (actual $\delta$ should be considerably smaller). We found that the remarkable enhancement of $T_c$ at an almost constant doping level in this superconductor is associated with the ordering of the apical oxygen (dopant).[16]

Since superconductivity takes place in the $CuO_2$ plane, the in-plane Cu-O bond length which is a measure of hybridization between $Cu3d_{x2-y2}$ and $O2p_{x,y}$ orbitals as well as of superexchange coupling between neighboring Cu spins will be correlated with the superconducting order parameters. The in-plane electronic parameters are also controlled by the charger reservoir layer, in particular, by the apical-O distance from the $CuO_2$ plane. Therefore, it is expected that $T_c$ will also be pertinent to the average size of A site ion. In fact, it has been found that the mean A-site cation size indeed correlates to superconducting transition temperature ($T_c$). For instance, in $La_{2-x}(Sr,Ba)_xCuO_4$ system with the optimum doping level, the variation of $T_c$ with A-site cation size has been investigated.[3] A systematic enhanced $T_c$ with increasing A

site ion size were observed.[17,18,19]

In fact, the out-plane effects induce the geometrical distortions of the unit cell. Nunner *et al.*[20] have proposed that the geometrical distortions of the unit cell may locally alter the electron-pairing interactions on micro scale level. These interactions are much related with the $T_c$ of the superconductors. In recent experiment, J. A. Slezak et al [21] had directly observed how variations in interatomic distances, presumably the apical-O distance, within individual crystalline unit cells affect the maximum superconducting energy-gap of $Bi_2Sr_2CaCu_2O_{8+\delta}$ (Bi-2212).

In the present study, we report the effect of average A-site cation size or the change of the lattice constants on $T_c$ for the $A_2CuO_{3+\delta}$ (A=Sr, Ba) superconductors. Since both Sr & Ba are isovalent alkaline earth element, their substitution does not change the doped carrier density when the excess oxygen content δ is kept constant. We have successfully synthesized a series of $Sr_{2-x}Ba_xCuO_{3+\delta}$ ($x \leq 0.6$) superconductors under high pressures and high temperatures. We found that $T_c$ can be significantly enhanced to 98 K in Ba-substituted $Sr_2CuO_{3+\delta}$.

## 2. Experimental

The samples were synthesized in two steps. In the first step, a series of $Sr_{2-x}Ba_xCuO_3$ ($0 \leq x \leq 0.6$) precursors were prepared by conventional solid-state reaction method from high-purity raw materials $SrCO_3$, $BaCO_3$ and CuO. The powder mixture in an appropriate ratio was ground together in an agate mortar and then calcined at 950°C for approximately three days in air with two intermediate grindings in order to obtain single-phase samples. In the second step, the precursors were mixed with $SrO_2$ and CuO to form the nominal composition of $Sr_{2-x}Ba_xCuO_{3+\delta}$ in a dry glove box that was used to prevent the degradation of hygroscopic reagents. The materials were then subjected to high pressure synthesis under 6 GPa pressure and at 1000°C for 30 min using a cubic-anvil-type high pressure facility, and then quenched to room temperature before releasing the pressure. The role of $SrO_2$ is to create an oxygen atmosphere during the high-pressure synthesis as previously described in the related Cl-series[22,23,24] or $Sr_2CuO_{3+\delta}$[16] superconductors. Here the excess oxygen content δ is controlled by the oxidizer amount $SrO_2$ in the nominal starting materials. In order to study the ordering effect of the dopant atoms (apical oxygen distribution), we also annealed several samples in the temperature range between 150 and 300°C for 12 h under 1 atm $N_2$ atmosphere in a tube furnace.

The structure of obtained polycrystalline samples was identified by powder x-ray diffraction (XRD) collected with 2θ-steps of 0.02° and 3 s counting time in the range of 5°≤ 2θ ≤ 120° using Cu Kα radiation. The x ray diffraction pattern was refined with Rietveld method using Winplotr program. The DC magnetic susceptibility was measured using a SQUID magnetometer in an external magnetic field of 20 Oe for both zero-field-cooling (ZFC) and field-cooling (FC) modes.

3. Results and discussion

By changing the Sr to Ba ratio and keeping a constant hole concentration (fixed δ), a series of $Sr_{2-x}Ba_xCuO_{3+\delta}$ (0≤ $x$ ≤ 0.6) superconductors were prepared using the high-pressure and high-temperature synthesis. Since a maximum $T_c$ is attained in SrCuO$_{3+\delta}$ at nominal $\delta$=0.4, we keep all the Sr$_{2-x}$Ba$_x$CuO$_{3+\delta}$ samples with the same oxygen content in order to separate out the A-site effects. The x-ray diffraction patterns show purely tetragonal K$_2$NiF$_4$-type phase with space group I4/mmm in the entire $x$ range up to 0.6. Note that no information on the apical-O configuration can be extracted directly from the x-ray diffraction. Assuming the La$_2$CuO$_4$ prototype tetragonal structure but with partially occupied apical oxygen, we refined the crystal structure of Sr$_{2-x}$Ba$_x$CuO$_{3+\delta}$ using the Rietveld method. Figure 2 shows the XRD pattern of a sample with the nominal composition Sr$_{1.84}$Ba$_{0.16}$CuO$_{3.4}$, together with the results of Rietveld refinement. A reasonable $R_{wp}$ (weighed-profile R-factor) factor of 8.76% was obtained as shown in Table 1, indicating that the structural model is a reasonable one. Figure 3 shows the cell parameters of the samples as a function of Ba content x for Sr$_{2-x}$Ba$_x$CuO$_{3+\delta}$. As compared with Sr$_2$CuO$_{3+\delta}$,[16] both $a$- and $c$-axis lattice constants of Sr$_{2-x}$Ba$_x$CuO$_{3+\delta}$ are expanded due to larger Ba ion incorporated to the Sr sites. For example, the lattice parameters increased from $a$=3.79 Å, $c$=12.50 Å for Sr$_2$CuO$_{3+\delta}$[16] to $a$= 3.89 Å, and $c$= 12.78 Å for Sr$_{1.4}$Ba$_{0.6}$CuO$_{3+\delta}$, respectively.

In Fig. 4(a) we present the temperature dependence of magnetic susceptibility measured for the as-prepared samples Sr$_{2-x}$Ba$_x$CuO$_{3+\delta}$ with x=0.16, 0.32, and 0.6 in the field-cooled (FC) mode at an external field of 20 Oe. The superconducting transition temperature $T_c$ increases with increasing Ba content. Starting from $T_c$=75K for Sr$_2$CuO$_{3+\delta}$ a maximum $T_c$ = 98 K is reached at x=0.6. The Meissner volume fraction of Sr$_{2-x}$Ba$_x$CuO$_{3+\delta}$ sample is more than 10% at 5 K, suggesting the nature of bulk superconductivity. The $Sr_{2-x}Ba_xCuO_{3+\delta}$ sample with $x>0.6$ are mixed phases, showing no superconducting transition.

In the case of $Sr_2CuO_{3+\delta}$, $T_c$ can be enhanced to 95K by post-annealing at relatively low temperatures (~250 °C), probably because the apical-O atoms order optimally. However, the Ba substituted samples very easily decomposes upon heating at ambient pressure. It is found that $Sr_{2-x}Ba_xCuO_{3+\delta}$ becomes unstable with increasing Ba content due to the fact that the high pressure synthesized materials with larger unit cell is usually thermodynamically metastable at ambient pressure. In Fig. 4(b) we show the post-annealing effect on $Sr_{1.9}Ba_{0.1}CuO_{3.4}$. $T_c$ increases from the initial 80K (as-prepared) to 91K by annealing at 250 °C, but its superconducting volume fraction appreciably decreases. With increasing Ba content to $x=0.32$, $T_c=88$K of as-prepared sample is enhanced to 92K by annealing at 200 °C, but annealing at higher temperatures results in degradation of superconductivity (Fig. 4(c)). Though the nominal composition is also $Sr_{1.68}Ba_{0.32}CuO_{3.4}$, the $T_c$ of the sample is a little lower than the $T_c$ of the sample in Fig.4(a). In fact, besides the nominal composition, the $T_c$ of the material varies a little depending on the synthesis temperature. In the case of $x=0.6$, as-prepared $T_c$ reaches 98K, already exceeding the maximum $T_c=95$K of $Sr_2CuO_{3+\delta}$. with 'optimally ordered apical-O atoms. We annealed the sample of $Sr_{1.4}Ba_{0.6}CuO_{3.4}$, but its superconductivity is lost immediately upon heating even at 150°C, indicating the heavily metastable nature of the Ba substituted sample. It seems that the highest $T_c$ of the present superconductor is realized in the vicinity of a structural instability.

For $Sr_2CuO_{3+\delta}$ the electron diffraction (ED) and high-resolution TEM measurements are possible to investigate the evolution of the structural modulation associated with the apical-O ordering with post-annealing temperature[16]. A modulated structure with C2/m symmetry is found for as-prepared sample with $T_c$=75K which evolves into a Pmmm modulation structure with $T_c$=95K after annealing at 250 °C. Supposing that the as-prepared samples have the same C2/m modulated structure as that of. $Sr_2CuO_{3+\delta}$, then the result indicates that $T_c$ is remarkably enhanced by Ba substitution without changing the ordering pattern of apical-O atoms. Unfortunately, the neutron or ED and TEM investigation are not possible, at the moment, for the Ba-substituted materials, since it is difficult to obtain sufficient amount of samples, and they are unstable against heating due to exposure to high density electron-beam flux. So, we cannot rule out the possibility that the Ba incorporation tends to stabilize the Pmmm modulation corresponding to the 'optimal' apical-O ordering in $Sr_2CuO_{3+\delta}$ with $T_c$=95K. In either case, since $T_c$=98K achieved in the as-prepared sample of x=0.6 exceeds the highest $T_c$ of $Sr_2CuO_{3+\delta}$, it is certain that the larger A-site cation size leads to higher $T_c$ in the present system.

It is also noted that the substituted Ba in the $Sr_2CuO_{3+\delta}$ plays dual roles; one is to expand the unit cell size, and the other is to create disorder due to cation size mismatch between $Sr^{2+}$ and $Ba^{2+}$. The previous investigation of $La_2CuO_4$-related superconductors suggested that $T_c$ depends both on the average cation size and the A-site cation disorder. It was also suggested that, if the degree of disorder does not change, $T_c$ tends to increase with increasing A-site cation size $<r_A>$, while $T_c$ decreases with increasing the A-site disorder at the constant average $<r_A>$. In the present $Sr_{2-x}Ba_xCuO_{3+\delta}$ system, $T_c$ of as-prepared sample increases with A-site cation size, achieving 98K at $x=0.6$. The result suggests that the *larger* A-site cation size is an important ingredient for the enhancement of $T_c$ in $Sr_2CuO_{3+\delta}$, while the A-site disorder arising from cation size mismatch between Sr and Ba appears to have relatively small effect on $T_c$. Since Sr and Ba are isovalent alkaline earth element, their substitution does not change the carrier density, nor introduce charge disorder, which is different from ordinary A-site substitution with hetero-valence ions such as that in $La^{3+}_{2-x}Sr^{2+}_xCuO_4$.[17,18,19] From the Shannon table [25], the ionic radius for $Ba^{2+}$ ($Sr^{2+}$) is 1.47Å (1.31Å). The calculated disorder parameter $\sigma^2 = <r_A^2>-<r_A>^2$ for $Sr_{2-x}Ba_xCuO_{3+\delta}$ is 0.005 for $x=0.6$. In the case of other single-layer cuprates, such as La-based 214 and rare-earth substituted Bi2201 compounds, this value is so large that it would have lead to a considerable reduction of $T_c$ – more than 50% reduction, or $T_c$ reduction of more than 30K, from the $T_c$ value of disorder-free ($\sigma=0$) material.[4] It is likely that a similar $T_c$-reduction took place in $Sr_2CuO_{3+\delta}$ by Ba-substitution, and an ideal $T_c$ value of 'disorder-free' $Sr_{1.4}Ba_{0.6}CuO_{3+\delta}$ would be 130K or higher. Note that in the La214 and Bi2201 systems the A-site disorder is accompanied with charge disorder. So, it is inferable that dominant effect of A-site disorder on $T_c$ is from charge disorder introduced by substitution of hetero-valent cations which is not the case with the present system.

If there were a small but finite reduction in $T_c$ due to the cation-size disorder, then the $T_c$-enhancement in $Sr_{2-x}Ba_xCuO_{3+\delta}$ is indicative of substantial contribution from the expansion of the unit cell size. There might be two factors associated with the increase of lattice parameters which would affect the electronic parameters in the $CuO_2$ plane; (1) increase in the apical-O distance, possibly caused by the increase of the c-axis lattice constant, and (2) increase in the Cu-O in-plane bond length. The correlation between $T_c$ and apical-O distance is empirically pointed out in the two contexts, a change in the 2$^{nd}$ (3$^{rd}$) nearest neighbor hopping parameter and a change in the Madelung energy. In a different context, Geballe suggests a possible role of apical-O atoms as negative-$U$ centers which promote pair formation[26].

The in-plane Cu-O bond length changes from 1.89Å for $Sr_2CuO_{3+\delta}$ to 1.95Å for $Sr_{2-x}Ba_xCuO_{3+\delta}$ at $x$=0.6. The simplest copper oxide containing $CuO_2$ plane is the so called infinite layer compound $CaCuO_2$ where in $CuO_2$ plane alternatively stacks with Ca spacer along the $c$ axis. The $CuO_2$ plane in the infinite layer compound is thus "free-standing" suffering only weak compressive or tensile stresses from the Ca layer. The Cu-O bond length in the infinite layer $CaCuO_2$ is 1.93Å [27] that is supposed to be the in-plane bond length of free-standing $CuO_2$ plane. In many materials the in-plane Cu-O bond length is appreciably shorter than 1.93Å, in which case the $CuO_2$ plane shows buckling. On the other hand, the bond length is close to or slightly longer than 1.93 Å in Tl[28]- and Hg[29]-based cuprates with $T_c$ higher than 90K. From this, it seems that *flat* $CuO_2$ plane with expanded dimension provides a stage favorable for higher $T_c$ as shown in Fig.5. In this sense, the $T_c$ enhancement with Ba substitution might be a consequence of the increasing Cu-O bonding length from 1.89 Å for $Sr_2CuO_{3+\delta}$ toward 1.93 Å. In the context a much higher Tc can be expected for $Ba_2CuO_{3+\delta}$.

## 4. Summary

In conclusion, we have successfully synthesized Ba-doped $Sr_{2-x}Ba_xCuO_{3+\delta}$ ($0 \leq x \leq 0.6$) with a $K_2NiF_4$-type structure using a high-pressure and high-temperature synthesis. The superconducting transition temperature with $T_c^{max}$ = 98 K is achieved for $x$=0.6. The increase in the unit cell size due to incorporation of large-size Ba cation is likely responsible for the enhancement of $T_c$ in this material. Since Ba is isovalent to Sr, the result also suggests that the disorder in the cation size introduced into the A-sites has minor effect on $T_c$. It is argued that the increase in the apical-O distance and/or the increase in the in-plane Cu-O bond length that makes the $CuO_2$ plane flat would favor higher $T_c$ in this material as well as other cuprates.

**Acknowledgment**: This work was supported by NSF & MOST of China through the research project, by a Grant-in-Aid for Scientific Research from MEXT, Japan, and a China-Korea-Japan A3 Forsite Program of the Japan Society for the Promotion of Science.

**Reference**

[1] H. Zhang, H. Sato, Phys. Rev. Lett. 70，1697 (1993).

[2] H. Eisaki, N. Kaneko, D. L. Feng, A. Damascelli, P. K. Mang, K. M. Shen, Z.-X. Shen, and M. Greven，Phys. Rev. B 69, 064512 (2004)

[3] J. P. Attfield, A. L. Kharlanov，and J. A. McAllister, Nature 394, 157 (1998).

[4] K. Fujita, T. Noda, K. M. Kojima, H. Eisaki, and S. Uchida, Physical Rev. Lett. 95，097006 (2005).

[5] A. Sugimoto, S. Kashiwaya, H. Eisaki, H. Kashiwaya, H. Tsuchiura, Y. Tanaka, K. Fujita, S. Uchida, Phys. Rev. B 74, 094503 (2006)

[6] K. McElroy, Jinho Lee, J. A. Slezak, D. –H. Lee, H. Eisaki, S. Uchida, J. C. Davis, Science 309, 1048 (2005).

[7] K. McElroy, R. W. Simmonds, J. E. Hoffman, D. –H. Lee, J. Orenstein, H. Eisaki, S. Uchida, J. C. Davis, Nature 422, 592 (2003).

[8] S. H. Pan, J. P. O'Neal, R. L. Badzey, C. Chamon, H. Ding, J. R. Engelbrecht, Z. Wang, H. Eisaki, S. Uchida, A. K. Gupta, K. –W. Ng, E. W. Hudson, K. M. Lang, J. C. Davis, Nature 413, 282 (2001).

[9] K. M. Lang, V. Madhavan, J. E. Hoffman, E. W. Hudson, H. Eisaki, S. Uchida, J. C. Davis, Nature 415, 412 (2002).

[10] G. Kinoda, T. Hasegawa, S. Nakao, T. Hanaguri, K. Kitazawa, K. Shimizu, J. Shimoyama, K. Kishio, Phys. Rev. B 67, 224509 (2003).

[11] Y. Kohsaka, K. Iwaya, S. Satow, T. Hanaguri, M. Azuma, M. Takano, H. Takagi,

Phys. Rev. Lett. 93, 097004 (2004).

[12]E. Pavarini, I. Dasgupta, T. Saha-Dasgupta, O. Jepsen, and O. K. Andersen, Phys. Rev. Lett. **87**, 047003 (2001).

[13]Z. Hiroi, M. Takano, M. Azuma and Y. Takeda, Nature **364**, 315 (1993).

[14]P. D. Han, L. Chang, D. A. Payne, Physica C **228**, 129 (1994).

[15] B. A. Scott, J. R. Kirtley, D. Walker, B. H. Chen, and Y. Wang, Nature **389**, 164 (1997).

[16] Q. Q. Liu, H. Yang, X. M. Qin, Y. Yu, L. X. Yang, F. Y. Li, R. C. Yu, and C. Q. Jin, Phys. Rev. B 74, 100506 R (2006).

[17]J. M. Tarascon, L. H. Greene, W. R. Mckinnon, G. W. Hull, Solid State Commun. 63, 4990505 (1987).

[18]K. Oh-Ishi, Y. Syono, Solid State Chem. 95, 136-144 (1991).

[19]B. Bűchner, M. Braden, M. Cramm, W. Schlabitz, O. Hoffels, W. Braunisch, R. Müller, G. Heger and D. Wohlleben, Physica C 185-189, 903-904 (1991).

[20]Tamara S. Nunner, Brian M. Andersen, Ashot Melikyan, and P. J. Hirschfeld, Phys. Rev. Lett. **95**, 177003 (2005)

[21]J. A. Slezak, Jinho Lee, M. Wang, K. McElroy, K. Fujita, B. M. Andersen, P. J. Hirschfeld, H. Eisaki, S. Uchida, and J. C. Davis, PANS 105,0706795105(2008)

[22]C. Q. Jin, X. J. Wu, P. Laffez, T. Tatsuki, T. Tamura, S. Adachi, H. Yamauchi, N. Koshizuka and S. Tanaka, Nature (London) 375 301 (1995).

[23]C. Q. Jin, X. J. Wu, S. Adachi, T. Tamura, T. Tatsuki, H. Yamauchi, S. Tanaka, and Z.-X. Zhao, Phys. Rev. B 61, 778 (2000); C. Q. Jin, High pressure research 24,

399 (2004).

[24] Q. Q. Liu, X. M. Qin, Y. Yu, F. Y. Li, C. Dong, C. Q. Jin, Physica C 420, 23 (2005).

[25] R. D. Shannon, Acta Cryst. A32, 751(1976).

[26] T. H. Geballe and Boris Y. Moyzhes, Physica C 341-348, 1821 (2000)

[27] T. Siegrist, S. M. Zahurak, D. W. Murphy, R. S. Roth, Nature 334 231(1988)]

[28] C.C. Torardi, M. A. Subramanian, J. C. Calabrese, J. Gopalakrishnan, E. M. McCarron, K. J. Morrissey, T. R. Askew, R. B. Flippen, U. Chowdhry, and A. W. Sleight, Phys. Rev. B 38, 225 (1988).

[29] S. N. Putilin, E. V. Antipov, O. Chmaissem, M. Marezio, Nature 362, 226 (1993).

[30] R. J Cava, A. Santoro, Jr D.W. Johnson, W. W Rhodes, Phys. Rev. B 35, 6716 (1987).

[31] J.D. Jorgensen, B. Dabrowski, S.Y. Pei, D.G. Hinks, L. Soderholm, B. Morosin, J. E. Schirber, E. L. Venturini, D. S. Ginley, Phys. Rev. B 38, 1337 (1988).

[32] D. C. Sinclair, S. Tait, J. T. Irvine, A. R. West, Physica C 205, 323 (1993)

[33] D. U. Gubser, R.A. Hein, S. H.Lawrence, M. S. Osofsky, D. J. Schrodt, L. E. Toth, and S. A.Wolf, Phys. Rev. B 35, 5350(1987)

# Table 1

Results of structure refinements for the superconductor $Sr_{1.84}Ba_{0.16}CuO_{3.4}$ using the Rietveld method. Space group: I4/mmm. $R_{wp}$=8.76%, $R_p$=6.15%. The lattice parameters: a= 3.8152(2)Å, c= 12.5665(3)Å. The refinement range of 2θ is 5 to 120°. CuKα1 radiation was used.

| Atom | Site | X | Y | Z | Occupancy |
| --- | --- | --- | --- | --- | --- |
| A(Sr/Ba) | 4e | 0 | 0 | 0.35590（14） | 1 |
| Cu | 2a | 0 | 0 | 0 | 1 |
| O1 | 4c | 0 | 0.5 | 0 | 1 |
| O2 | 4e | 0 | 0 | 0.172072(0) | 0.58(2) |

# Figure captions

Fig.1 The crystal structure of (Sr Ba)$_2$CuO$_{3+\delta}$, similar to La$_2$CuO$_4$ but with partially occupied apical oxygen sites.

Fig.2. The Rietveld refinement results of powder XRD patterns for Sr$_{1.84}$Ba$_{0.16}$CuO$_{3.4}$ sample. Dots and lines show the observed and calculated patterns, respectively. The difference between the observed and fitted patterns is displayed at the bottom of the figures.

Fig.3. The lattice parameters *a* and *c* as a function of x value for Sr$_{2-x}$Ba$_x$CuO$_{3+\delta}$.

Fig.4. (a).Temperature dependence of the dc magnetic susceptibility in the field-cooling mode for the samples prepared at high pressure and high temperature from nominal compositions of Sr$_{2-x}$Ba$_x$CuO$_{3.4}$ with various x values. (b). Temperature dependence of the dc magnetic susceptibility in the field-cooling mode for as-prepared Sr$_{1.9}$Ba$_{0.1}$CuO$_{3.4}$ sample and those after annealed at various temperatures in the N$_2$ atmosphere. (c) Temperature dependence of the dc magnetic susceptibility in the field-cooling mode for another as-prepared Sr$_{1.68}$Ba$_{0.32}$CuO$_{3.4}$ and those after annealed at various temperatures in the N$_2$ atmosphere.

Fig.5. The T$_c$ dependence on the in plane CuO bond length for various single layered HTS at an optimal doping level {Ref: La$_{1.85}$Sr$_{0.15}$CuO$_4$ [30]; La$_2$CuO$_{4+\delta}$ [31]; Bi$_2$Sr$_2$CuO$_8$ [32], La$_{1.85}$Ba$_{0.15}$CuO$_4$ [33]; La$_{1.8}$Ca$_{0.2}$CuO$_4$ [33]; Tl$_2$Ba$_2$CuO$_5$ [28]; HgBa$_2$CuO$_{4+\delta}$ [29]; Sr$_{1.4}$Ba$_{0.6}$CuO$_{3+\delta}$ [this work]}.

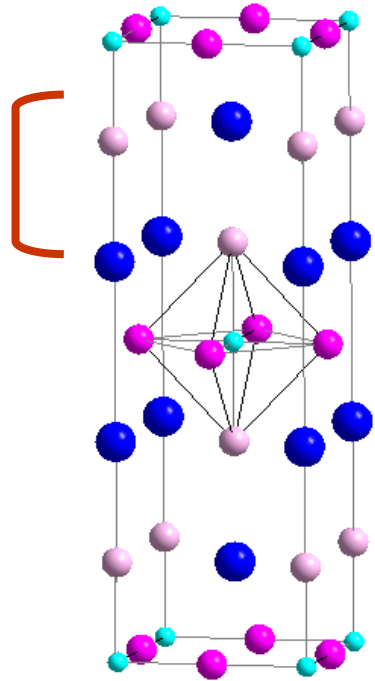

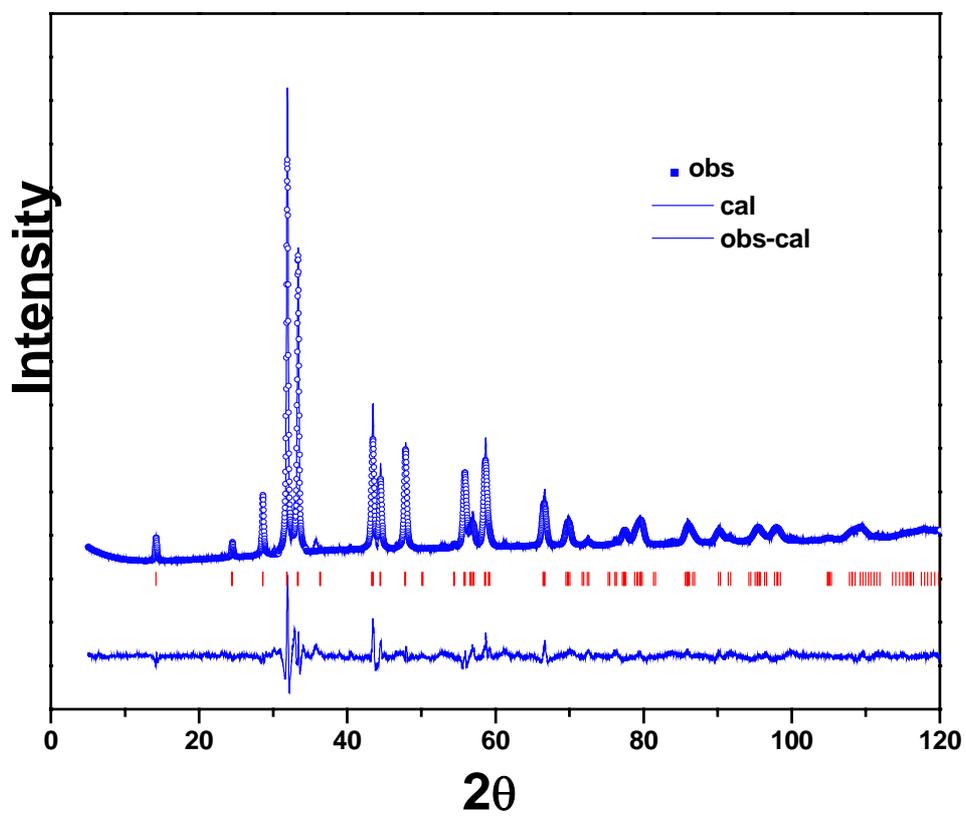

Gao *et al.*

Fig.2

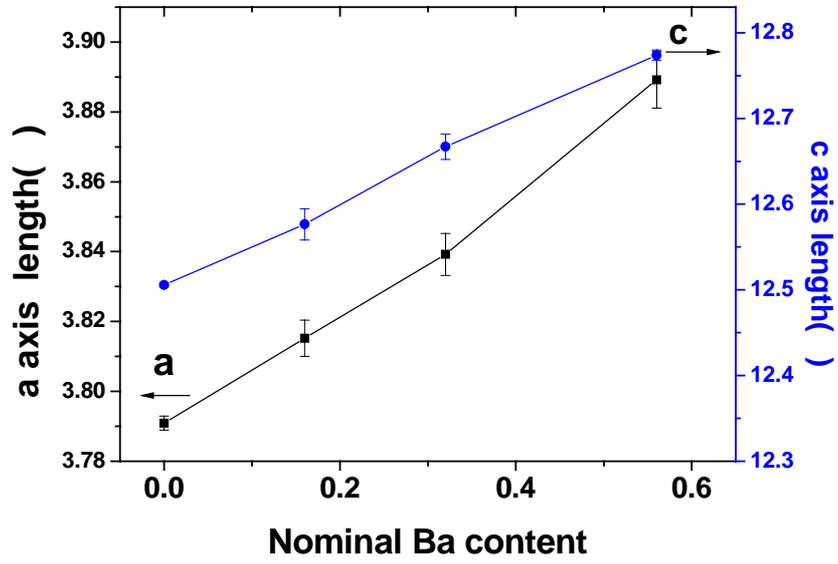

Gao *et al.*

Fig.3

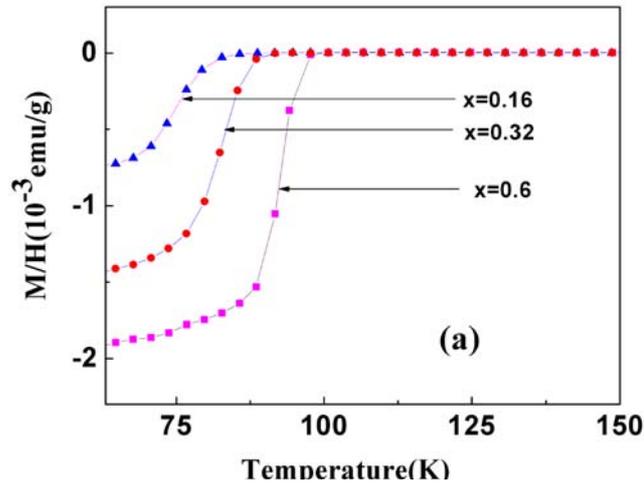
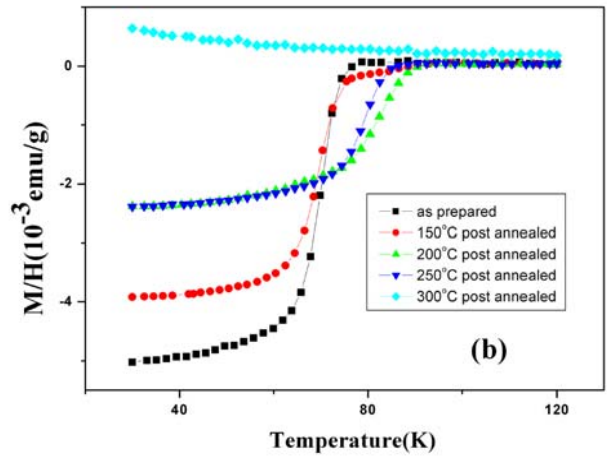
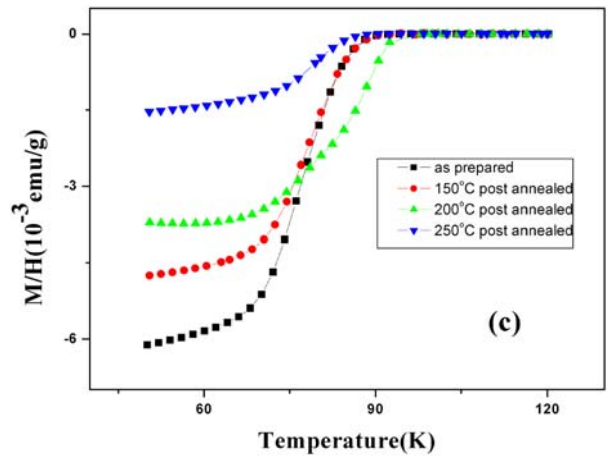

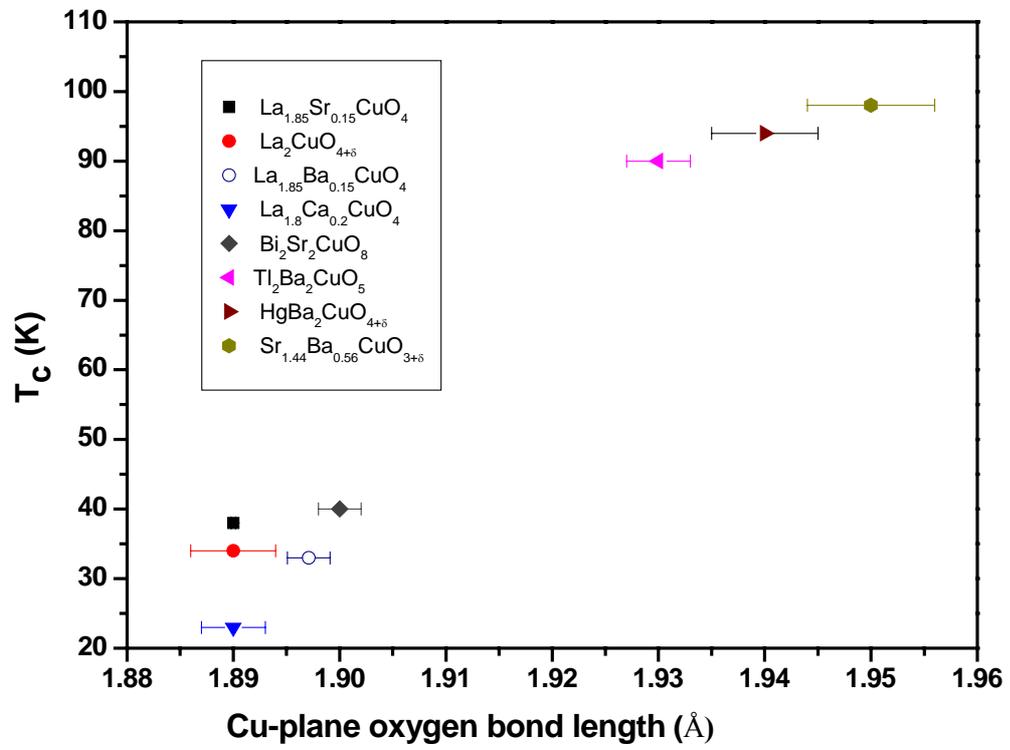

Gao *et al.*

Fig.5